\documentclass{emulateapj} 
\usepackage{graphicx}
\usepackage{amsmath}
\begin{document}
%
\title{Interacting quark matter equation of state for compact stars}
\hfill CERN-PH-TH/2013-269, HIP-2013-27/TH

\author{Eduardo S. Fraga$^{1,2,3}$, Aleksi Kurkela$^4$ and Aleksi Vuorinen$^5$}
\affil{$^1$Institute for Theoretical Physics, Goethe University, D-60438 Frankfurt am Main, Germany}
\affil{$^2$Frankfurt Institute for Advanced Studies, Goethe University, D-60438 Frankfurt am Main, Germany}
\affil{$^3$Instituto de F\'\i sica, Universidade Federal do Rio de Janeiro,
Caixa Postal 68528, 21941-972, Rio de Janeiro, RJ, Brazil}
\affil{$^4$Theory Division, PH-TH, Case C01600, CERN, CH-1211 Geneva 23, Switzerland}
\affil{$^5$Department of Physics and Helsinki Institute of Physics, P.~O.~Box 64, FI-00014 University of Helsinki, Finland}

\begin{abstract}
Lattice QCD studies of thermodynamics of hot quark-gluon plasma (QGP) demonstrate the importance of accounting for the
interactions of quarks and gluons, if one wants to investigate the phase structure of strongly interacting matter. Motivated by this observation and using state-of-the-art results from perturbative QCD, we construct a simple effective equation of state for cold quark matter that consistently incorporates the effects of interactions and furthermore includes a built-in estimate of the inherent systematic uncertainties. This goes beyond the MIT bag model description in a crucial way, yet leads to an equation of state that is equally straightforward to use. We also demonstrate that, at moderate densities our EoS can be made to smoothly connect to hadronic ones, with the two exhibiting very similar behavior near the matching region. The resulting hybrid stars are seen to have masses similar to those predicted by the purely nucleonic EoSs.
\end{abstract}


\preprint{CERN-PH-TH/2013-269,HIP-2013-27/TH}

\section{Introduction}
\label{sec:intro}

The structure of a stellar object is determined by a competition between its internal pressure and gravity, with the former given by the equation of state (EoS) of the matter composing the star \citep{Glendenning:1997wn}. For compact stars, such as (hybrid) neutron or quark stars, what is required is the EoS of strongly interacting matter at high density and low temperature. Such information is in principle available from the underlying theory, Quantum Chromodynamics (QCD), which, however, has so far not admitted a nonperturbative first principles solution at nonzero quark (or baryon) density. The reason for this is the well-known Sign Problem of lattice QCD \citep{deForcrand:2010ys}.

At low densities, the physics determining the stellar EoS is that of low-energy nuclear interactions, which can be treated successfully through relativistic or non-relativistic effective theories (see \cite{Lattimer:2012nd} and references therein). In the cores of the stars, the density of the matter is, however, likely to significantly exceed that of nuclear saturation density. This may lead to the presence of matter better described by deconfined quark than hadronic degrees of freedom. In this region, a low-energy description of the thermodynamics indeed clearly breaks down, leaving one with very limited machinery for the determination of the EoS. In the absence of more realistic competitors, the compact star community has widely adopted the simplest alternative, the MIT bag model EoS \citep{Farhi:1984qu}, in which all effects from interactions are contained in one additive `bag' constant arising from the energy difference between the perturbative ground state and the physical, chiral symmetry breaking vacuum of the theory. This model was used in the very first works on quark stars \citep{Itoh:1970uw,Witten:1984rs,Alcock:1986hz,Haensel:1986qb}, and has since then been employed in most investigations of hybrid and quark stars as well as core collapse supernovae
\citep{Berezhiani:2002ks,Alford:2004pf,Drago:2005yj,Alford:2006vz,Mintz:2009ay,Fischer:2010wp,Ozel:2010bz,Weissenborn:2011qu,Zdunik:2012dj,Drago:2013fsa}. For some exceptions, using different low-energy models or the quasiparticle approach, see e.g. \cite{Dexheimer:2012eu,Klahn:2013kga,Orsaria:2013hna,Peshier:2002ww}.

In addition to its simplicity, the MIT bag model owes its popularity to its success in describing the vacuum properties of hadrons --- despite only incorporating confinement
in the crudest possible fashion. In the context of high temperatures, it has, however, been long known that the bag model fails badly in the description of thermodynamics:
it assumes that interactions can be altogether neglected above the phase transition region, a claim that has been convincingly disproven by lattice QCD simulations. Indeed, at any phenomenologically relevant temperature, the strong coupling constant $\alpha_s(T) \sim 1/\log(T/\Lambda_\text{QCD})$ is not vanishingly small, and perturbative corrections proportional to its powers (and logarithms) are numerically significant.

In the regime of high density and low temperature, common wisdom based on the MIT bag model EoS states that the masses of pure quark stars are unlikely to reach $2M_{\odot}$, and hybrid stars containing quark matter cores are generically less massive than their purely nucleonic counterparts \citep{Weissenborn:2011qu}. This conclusion was, however, challenged more than ten years ago in \cite{Fraga:2001id,Fraga:2001xc}, where a study of the perturbative EoS of high density quark matter was seen to lead to quark stars with masses above $2M_{\odot}$.\footnote{A comprehensive phenomenological study using this parametrization of the EoS, including effects from color superconductivity and matching onto different low-density nuclear EoSs, was later performed by \cite{Alford:2004pf}.} Somewhat later, the three-loop perturbative EoS used in these works was further extended by including the effects of the strange quark mass, leading to quark stars with masses in excess of $2.5M\odot$ \citep{Kurkela:2009gj} (c.f. \cite{Fraga:2004gz} for a discussion of quark mass effects at two loops). At that time, these masses were considered unrealistically high but since then the discoveries of the pulsars PSR J1614-2230, with $M = 1.97 \pm 0.04 M\odot$ \citep{Demorest:2010bx}, and PSR J0348+0432, with $M = 2.01 \pm 0.04 M\odot$ \citep{Antoniadis:2013pzd}, have dramatically changed the situation.

In the letter at hand, our aim is to package the state-of-the-art perturbative QCD EoS of \cite{Kurkela:2009gj} in a simple, easy-to-use form that can be used in modeling cold quark matter in compact stars. Although the diagrammatic computation behind this result is very involved, we have observed that it is possible to condense the EoS into a simple analytic fitting function that provides an accurate description of the pressure and its two first derivatives with respect to the baryon number chemical potential. As we will explicitly demonstrate in the following, this result is immediately amenable to the determination of the structure of quark stars, and can in addition be matched to various nuclear EoSs with considerable ease.

As our result constitutes a purely diagrammatic correction to the pressure of free quarks, it does not incorporate any fundamentally non-perturbative contributions to the EoS. While we believe that the breakdown of the perturbative series around nuclear matter saturation density signals the presence of non-perturbative physics, a calculation of this type cannot provide a quantitive description of these effects. Nevertheless, it is still possible to use our result to improve the perturbative part of the EoS in calculations including non-perturbative physics, such as a bag constant\footnote{In the MIT bag model, confinement is modeled by adding to the free pressure a ``bag constant'', corresponding to the free energy difference of the perturbative and non-perturbative, chiral symmetry breaking vacua. It is not clear that this physical effect can be included in the EoS in the form of one additive constant, but it is also no less consistent to add such a term to our EoS than to the noninteracting one.} or the color superconducting gap.

While being as convenient to use as the MIT bag model EoS, the result we provide improves it in two important ways. First, as demonstrated in the case of high temperature and vanishing density in Section \ref{sec:thermal}, the MIT bag model crudely underestimates the importance of interactions at large and intermediate energies, thus leading to an inaccurate description of the EoS. Secondly --- and perhaps more importantly --- our perturbative EoS automatically includes an estimate of its inherent systematic uncertainties via a dependence on the renormalization scale parameter. Thus, it replaces the fallaciously single-valued prediction of the MIT bag model EoS by a band, the width of which correctly increases at low densities, where the uncertainty of any quark matter EoS is bound to be sizable.

\section{The quark-gluon plasma equation of state}
\label{sec:thermal}

To motivate our following discussion of the EoS of cold quark matter, we first discuss the analogous problem at high temperature and vanishing baryon density. Here, one is faced with the task of connecting a low-energy hadron resonance gas EoS to the high-temperature limit, where a perturbative description of the thermodynamics is expected to be feasible. At intermediate temperatures, one has the options of using resummed perturbation theory, the MIT bag model, or other, more elaborate model calculations to estimate the behavior of the EoS. The matching of these EoSs to the hadron resonance gas results can in addition be used to determine the approximate location of the crossover transition, where the degrees of freedom used to describe the system effectively change from hadronic to partonic ones.

\begin{figure}
\includegraphics[width=8cm]{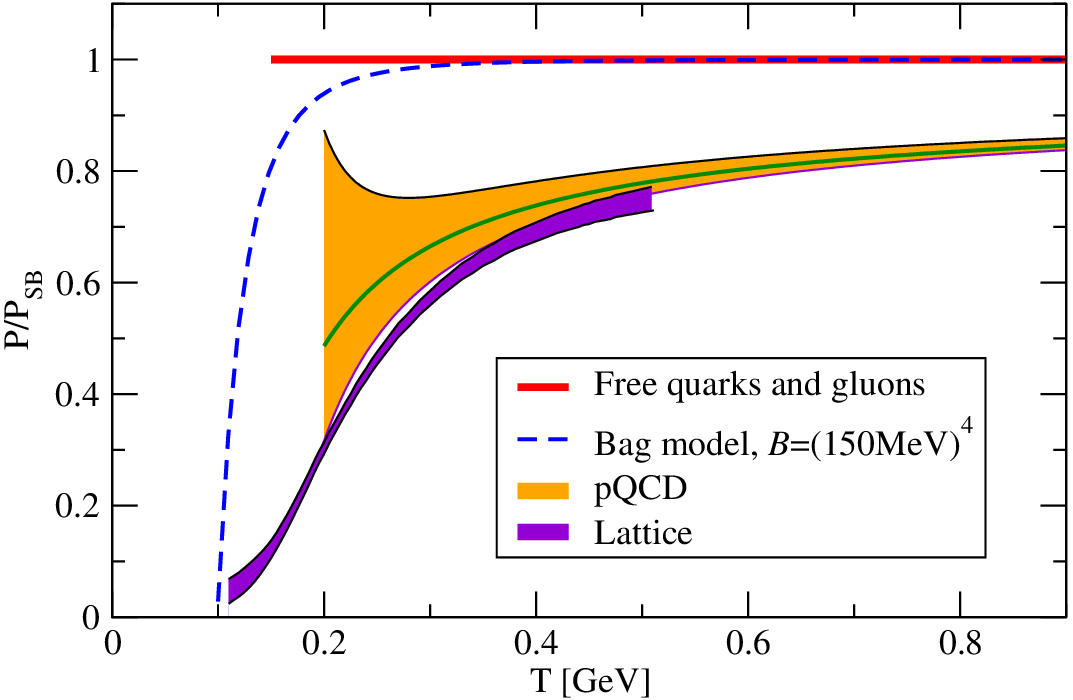}
\caption{Predictions for pressure of hot three-flavor QGP obtained from lattice QCD, the MIT bag model and perturbative QCD. The error bars reflect  various uncertainties in the results. The quantities are normalized by the pressure of a system of free quarks and gluons.}
\label{fig:eos-thermal}
\end{figure}

Despite the obvious similarities of the two systems, there is one crucial difference between the description of QCD matter at high density and high temperature: for hot and dilute QGP, lattice QCD provides a reliable nonperturbative first principles method to evaluate bulk thermodynamic quantities, and thus check the accuracy of the perturbative and model predictions. This is illustrated in Fig.~\ref{fig:eos-thermal}, where we plot the pressure of hot QGP at vanishing baryon number density, normalized to that of a free system. The figure exhibits a perturbative band obtained from the so-called `dimensional reduction resummation' (see
\cite{Blaizot:2003iq,Laine:2006cp}) of the highest complete weak-coupling result of order $\alpha_{s}^{5/2}$ \citep{Zhai:1995ac,Kajantie:2002wa},\footnote{Note that at high temperatures the strong infrared $(T/p)$-tail of the gluonic Bose-Einstein distribution renders the EoS sensitive to the dispersion relation of soft gluons (with $p\sim \alpha_s^{1/2}T$), creating a need for performing resummations. This complication is largely absent in the case of cold fermionic matter.} as well as a curve corresponding to the prediction of the MIT bag model, in which we have used the typical value of $(150~{\rm MeV})^4$ for the bag constant. The two predictions are compared to state-of-the-art lattice data from \cite{Borsanyi:2010cj, Borsanyi:2013bia}, corresponding to two massless and one massive flavor of dynamical quarks.

What one sees in Fig.~\ref{fig:eos-thermal} is quite striking: whereas the resummed perturbative EoS clearly provides a successful description of the zero density EoS (and even more so at small but nonzero baryon number chemical potential $\mu_B$ \citep{Mogliacci:2013mca}), the MIT bag model prediction fails badly in the same task. Inspecting the figure, one quickly sees that this can be attributed to the extremely sharp rise of the bag curve towards the noninteracting Stefan-Boltzmann limit, proportional to $1/T^4$. At the same time, the logarithmic dependence of the perturbative EoS on the temperature, originating from the running of the strong coupling constant, is in excellent agreement with the slow rising of the lattice data.

As noted above, at nonzero baryon density there is at the moment still no benchmark from lattice QCD, to which one could compare perturbative and model results. At the same time, there is, however, also no physical reason to expect interactions to play any smaller of a role in this regime, or that the bag model would provide a better description of the EoS. Therefore, if one wishes to allow for the presence of quark matter inside compact stars, it is crucial to use an EoS that correctly accounts for the interactions. As we will argue in the following section, such a result is in fact already available, and is provided by state-of-the-art perturbative QCD.

\section{The high-density EoS: Beyond the bag}

\label{sec:beyond}

Within perturbative thermal field theory, the EoS of a given system is obtained by expanding the path integral representation of the partition function in terms of zero-point Feynman diagrams. The expansion is, however, somewhat complicated by the fact that diagrams with any number of loops can contribute at the same order in $\alpha_s$. This is seen explicitly in the fact that at order $\alpha_s^2$ the pressure of zero-temperature QCD obtains contributions from an infinite set of so-called plasmon or ring diagrams \citep{Kraemmer:2003gd}.

Having determined the weak coupling expansion to a given order in $\alpha_s$, we observe that the result has become a function of an unphysical auxiliary parameter, the \emph{renormalization scale} $\bar{\Lambda}$. As long as the perturbative expansion converges, this dependence is, however, guaranteed to decrease order by order, and thus the sensitivity of our result on the parameter can be interpreted as reflecting the systematic error introduced by the truncation of the series. This error is commonly estimated by choosing a physically reasonable fiducial scale and varying the renormalization scale around it by a factor of two; below, we too follow this procedure, choosing as the central scale the commonly used value $\bar{\Lambda} = (2/3)\mu_B$.

For the pressure of QCD at nonzero density, the weak coupling expansion has so far been determined to ${\mathcal O}(\alpha_{s}^{3}\ln\,\alpha_{s})$ at temperatures $T\gtrsim \sqrt{\alpha_{s}} \mu$ \citep{Vuorinen:2003fs,Ipp:2006ij}, to ${\mathcal O}(\alpha_{s}^2)$ at $T=0$ \citep{Freedman:1976ub,Baluni:1977ms,Blaizot:2000fc,Fraga:2001id,Kurkela:2009gj}, and to ${\mathcal O}(\alpha_{s}^2\ln\, \alpha_{s})$ between these two limits \citep{Toimela:1984xy} (see also \cite{Andersen:2002jz,Gerhold:2004tb}). The calculation relevant for compact star physics is clearly the $\mathcal{O}(\alpha_s^2)$ zero-temperature work of \cite{Kurkela:2009gj}, which most importantly also takes into account the nonzero value of the strange quark mass. It is exactly this EoS, applied to the special case of electrically neutral and $\beta$-stable quark matter, that we will analyze in the present letter.\footnote{There is some freedom involved with the choice of the thermodynamical potential that one chooses to truncate at a given order in $\alpha_s$, while other functions are derived from it demanding thermodynamic consistency. Unlike in \cite{Kurkela:2009gj}, we have for simplicity chosen to truncate here the pressure as a function of $\mu_B$.} It is a function of the baryon chemical potential $\mu_B$ and parametrized by the strong coupling constant and strange quark mass, which are taken at arbitrary reference scales, $\alpha_s(1.5 {\rm GeV}) = 0.326$ and $m_s(2 {\rm GeV})=0.938\rm GeV$ \citep{Bazavov:2012ka,Aoki:2013ldr}, and then let evolve as functions of the $\overline{\rm MS}$ scale $\bar{\Lambda}$.
\begin{figure}
\includegraphics[width=8cm]{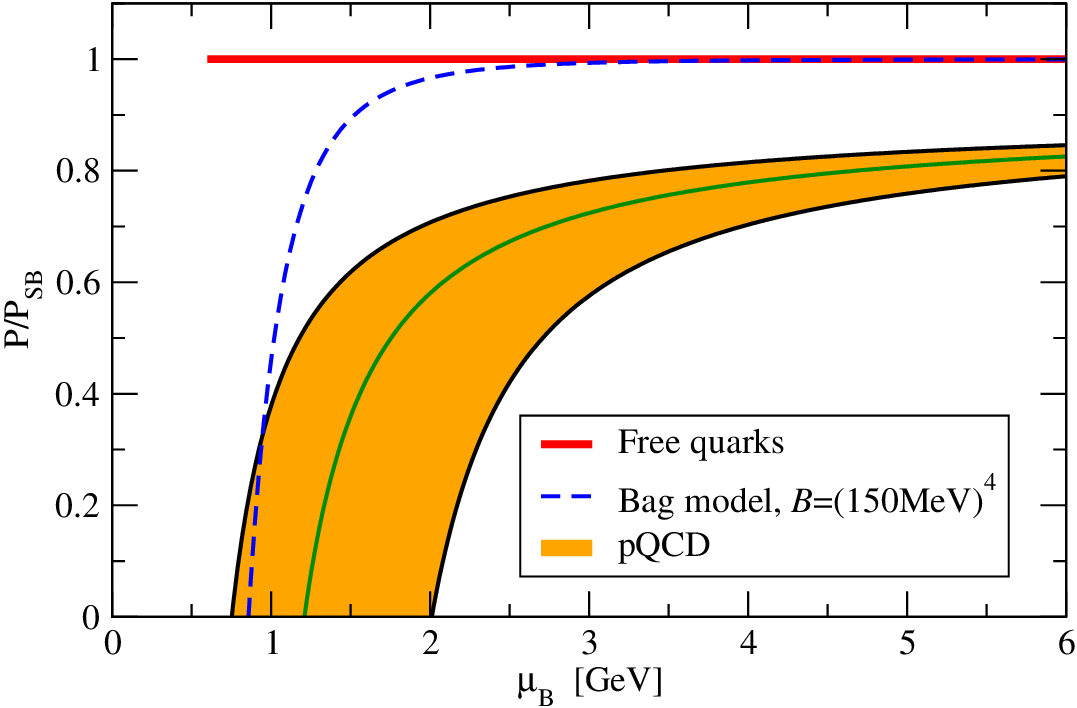}
\caption{Same as in Fig.~\ref{fig:eos-thermal}, but for the pressure of zero-temperature quark matter in $\beta$ equilibrium as a function of the baryon chemical potential.}
\label{eos-dense}
\end{figure}

We find that the EoS and its first and second derivatives are to a very good accuracy described by the compact fitting function
\begin{align}
P_{\rm{QCD}}(\mu_B,X) &= P_{\rm{SB}}(\mu_B) Ê\left( c_1 - \frac{a(X)}{(\mu_B/{\rm GeV}) - b(X)} Ê\right), \label{eq:pressure}\\
a(X) &= d_1 X^{-\nu_1},\quad
b(X) = d_2 X^{-\nu_2},
\end{align}
where we have denoted the pressure of three massless noninteracting quark flavors (at $N_c=3$) by
\begin{align}
P_{\rm SB}(\mu_B) = \frac{3}{4\pi^2}(\mu_B/3)^4.
\end{align}
The dependence of the result on the renormalization scale is contained in the functions $a(X)$ and $b(X)$, which depend on a dimensionless parameter proportional to the scale parameter, $X\equiv 3\bar{\Lambda}/\mu_B$, that is allowed to vary from $1$ to $4$.

The values of the constants $\{ c_1,d_1,d_2,\nu_1,\nu_2\}$ are fixed by minimizing the value of the following merit function
\begin{align}
\chi^2=[\Delta P(\mu_B,X)]^2 + [\Delta N(\mu_B,X)]^2 + [\Delta c_s^2(\mu_B,X)]^2,
\end{align}
where $\Delta P$, $\Delta N$, and $\Delta c_s^2$ are the differences between the values of the pressure, quark number density and speed of sound squared obtained from the fit and from the corresponding full perturbative expressions of \cite{Kurkela:2009gj}, normalized to the corresponding Stefan-Boltzmann values. For our best fit values
\begin{eqnarray}
c_1=0.9008 \quad & d_1= 0.5034 &\quad Êd_2 = 1.452 \\
\nu_1 &= 0.3553 \quad \nu_2&= 0.9101,
\end{eqnarray}
we obtain a good fit ($\sqrt{\chi^2}\lesssim 0.03$) in the region defined by the conditions $\mu_B< 2\rm{GeV}$, $P(\mu_B)>0$, and $X\in[1,4]$. We have checked that all relevant observables depending on the pressure and its first and second derivatives (such as the energy density as a function of pressure) are faithfully described by the fit.

To investigate the dependence of our result on the strange quark mass, we have also performed fits at unphysical quark masses, varying $m_s(2\rm{GeV})$ between 0 and $140\rm{MeV}$. The authors of \cite{Alford:2004pf} argue that the inclusion of a nonzero strange quark mass should generate a term in the EoS scaling with $\mu_B^2$. In the interacting case, involving a running quark mass, things are however significantly more complicated, and the dependence of our result on $m_s(2 {\rm GeV})$ is not well reproduced by a simple $\mu_B^2$ term. Rather, we find that the dependence is well described by rescaling the parameter $\mu_B$ in the EoS: If we write
\begin{align}
 (2 - X m_2/{\rm GeV})\mu_B' = Ê(2- X m_1/{\rm GeV})\mu_B,
\end{align}
then the simple relation $P(m_1, \mu_B)/P_{\rm SB}(\mu_B) = P(m_2, \mu_B')/P_{\rm SB}(\mu_B')$ is seen to hold within the accuracy of the fitting function in the entire range of parameters considered above. Here, the $m_i$ again stand for the strange quark masses measured at the scale $2\rm{GeV}$.

Finally, it should be noted that during the past 15 years there have been several attempts to describe various perturbative EoSs with fitting functions of increasing sophistication (see e.g.~\cite{Fraga:2001id, Alford:2004pf}). For the sake of comparison, we have also analyzed these \textit{ans\"atze}, and have found that our fitting function consistently provides $\chi^2$-values at least 40 times smaller than any of its competitors.

\begin{figure}
\includegraphics[width=8cm]{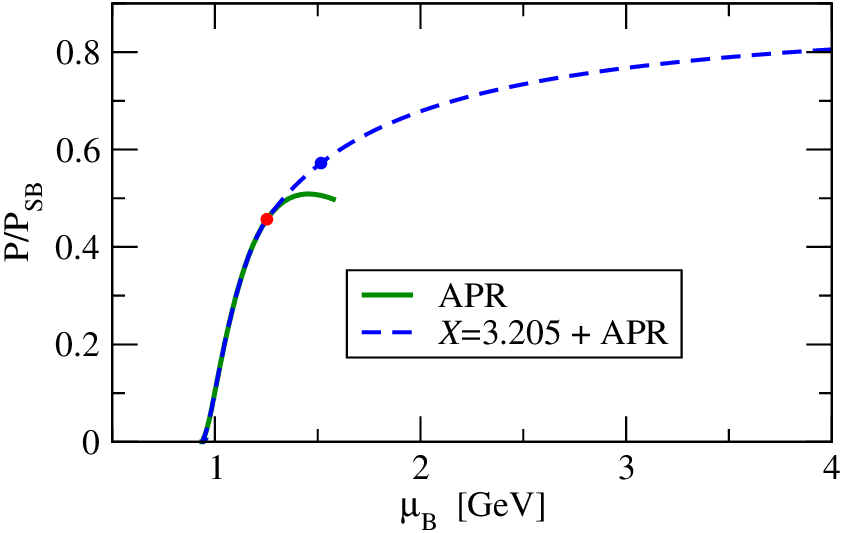}
\caption{The pressure resulting from the matching of our EoS with that of \cite{Akmal:1998cf}. Note that the behaviors of the APR and quark matter EoSs are strikingly similar in the region near the matching point, marked here by the red dot.}
\label{fig:eos-matched}
\end{figure}

\begin{figure}
\includegraphics[width=8cm]{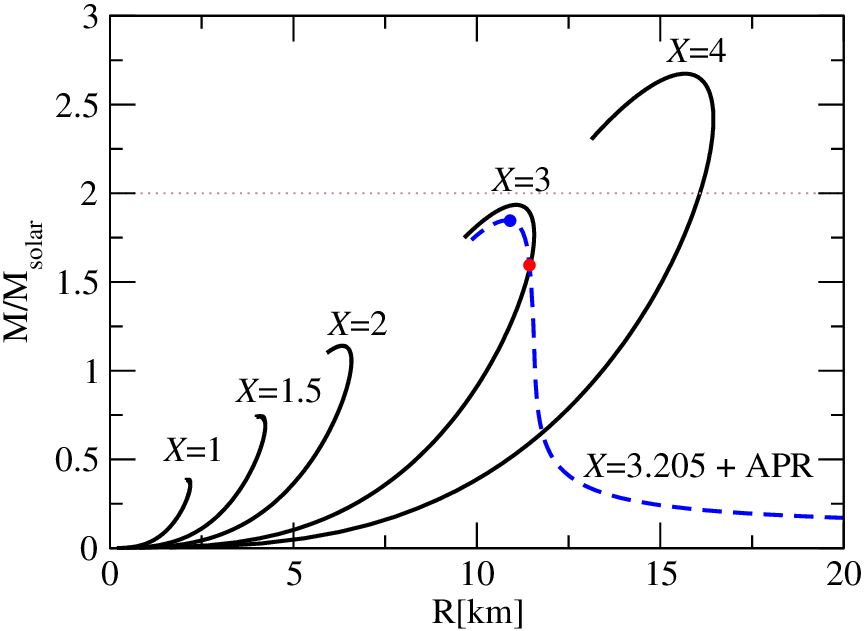}
\caption{A mass-radius diagram for pure quark and hybrid quark-neutron stars generated using the equation of state of Eq.~(\ref{eq:pressure}) as well as that displayed in Fig.~\ref{fig:eos-matched}, respectively. 
The red and blue dots on the hybrid curve stand for the cases, where the maximal density inside the star reaches the corresponding two points in Fig.~3.
 }
\label{fig:MR-diagram}
\end{figure}

To demonstrate the uses of our effective equation of state, we next determine the mass-radius relations of pure quark stars as well as a hybrid star obtained by joining our EoS to the standard nucleonic EoS (APR) of \cite{Akmal:1998cf} via the Maxwell construction (cf.~Fig.~\ref{fig:eos-matched} for the resulting pressure). The latter calculation is performed in the extremal case, where the latent heat vanishes and the pressure is continuous at the transition,\footnote{This takes place at $X=3.205$, for which the critical chemical potential, at which the quark and nuclear matter pressures match, reads $\mu_c = 1.27 {\rm GeV}$ and the pressure takes the value $P(\mu_c)=0.15{\rm GeV}/{ \rm fm}^3 $. The baryon number density at this point (in both phases, as in this extremal case where also the first derivatives of the pressure coincide) is $N(\mu_c) = 0.64/{\rm fm}^3$.} and is meant only as an illustration of the fact that such matchings are highly straightforward to implement. The resulting mass-radius curves, obtained by solving the Tolman-Oppenheimer-Volkov (TOV) equations \citep{Glendenning:1997wn}, are displayed in Fig.~\ref{fig:MR-diagram}. From here, we see that the maximal masses of the quark stars depend strongly on the value of the renormalization scale, and that it is not difficult to reach masses in excess of $2M\odot$.

\section{Conclusions}

In spite of its rudimentary nature, the MIT bag model has for decades been the framework of choice in studies of quark matter in compact stars. Primary questions, such as the possible existence of deconfined matter in the cores of neutron stars, have thus been addressed essentially neglecting all interactions between the constituents of the system. In light of recent, important advances in perturbative QCD at high density, it is clear that the state-of-the-art weak coupling EoS for cold quark matter should become the new standard in these investigations.

For the above reasons, the letter at hand has been aimed at providing a simple parametrized EoS for cold quark matter, which significantly improves the picture provided by the bag model (and other low-energy effective models). Most importantly, it is based on a first principles calculation within the fundamental microscopic theory \citep{Kurkela:2009gj}, and automatically contains a realistic estimate for the magnitude of its inherent theoretical uncertainties. At the same time, it is, however, as simple to use as the MIT bag model EoS, having been condensed to a single-line analytic formula in Eq.~(\ref{eq:pressure}). To demonstrate the impact this result has on compact star physics, we determined the mass-radius relation of pure quark stars based on the new EoS. In glaring contrast to studies carried out with the bag model EoS, this showed that it is in fact easy to construct pure quark stars with masses in excess of $2 M_{\odot}$, consistent with current observational bounds.

Finally, one should note that in the present work we only briefly experimented with the matching of our EoS to its low-density hadronic counterparts. A naive matching to the APR result demonstrated a striking similarity between the two EoSs near the transition point, and lead to stars almost as massive as purely nucleonic ones. It is certainly possible to carry out this exercise in a much more systematic way, which we in fact aim to do in a separate publication (see also \cite{Kurkela:2010yk}). There, one may additionally consider including in the EoS various nonperturbative contributions originating e.g.~from quark pairing or nontrivial vacuum properties (cf.~our discussion of the bag constant in footnote 3). After all, our current result is simply the \textit{perturbative EoS of unpaired quark matter}, aimed at replacing the free quark matter part in the MIT bag model EoS.

\textit{Acknowledgments}
The authors thank Zolt\'an Fodor, Paul Romatschke, J\"urgen Schaffner-Bielich, and Michael Strickland for useful discussions. In addition, ESF and AK thank the Helsinki Institute of Physics, and ESF and AV the EMMI Rapid Reaction Task Force: \textit{Quark Matter in Compact Stars} for hospitality. The work of ESF was financially supported by the Helmholtz International Center for FAIR within the framework of the LOEWE program (Landesoffensive zur Entwicklung Wissenschaftlich-\"Okonomischer Exzellenz) launched by the State of Hesse, while AV was supported by the Academy of Finland, grant \# 266185.\\

\end{document}